\documentclass[%
 reprint,
 amsmath,amssymb,
 aps,
]{revtex4-1}
\usepackage{changes} 
\usepackage{graphicx}
\usepackage{dcolumn}
\usepackage{bm}

\begin{document}


\title{Gate-tunable near-field heat transfer}

\author{Georgia T. Papadakis}
\affiliation{Department of Electrical Engineering, Ginzton Laboratory, Stanford University, California, 94305, USA}
\author{Bo Zhao}
\affiliation{Department of Electrical Engineering, Ginzton Laboratory, Stanford University, California, 94305, USA}
\author{Siddharth Buddhiraju}
\affiliation{Department of Electrical Engineering, Ginzton Laboratory, Stanford University, California, 94305, USA}
\author{Shanhui Fan}\email{shanhui@stanford.edu}
\affiliation{Department of Electrical Engineering, Ginzton Laboratory, Stanford University, California, 94305, USA}

\date{\today}

\begin{abstract}
Active control over the flow of heat in the near-field holds promise for nanoscale thermal management, with applications in refrigeration, thermophotovoltaics, and thermal circuitry. Analogously to its electronic counterpart, the metal-oxide-semiconductor (MOS) capacitor, we propose a thermal switching mechanism based on accumulation and depletion of charge carriers in an ultra-thin plasmonic film, via application of external bias. In our proposed configuration, the plasmonic film is placed on top of a polaritonic dielectric material that provides a surface phonon polariton (SPhP) thermal channel, while also ensuring electrical insulation for application of large electric fields. The variation of carrier density in the plasmonic film enables the control of the surface plasmon polariton (SPP) thermal channel. We show that the interaction of the SPP with the SPhP significantly enhances the net heat transfer. We study SiC as the oxide and explore three classes of gate-tunable plasmonic materials: transparent conductive oxides, doped semiconductors, and graphene, and theoretically predict contrast ratios as high as $225\%$.
\end{abstract}

\pacs{Valid PACS appear here}
\maketitle


\section{\label{Intro}INTRODUCTION}

\par{Controlling the flow of an electric current is the cornerstone of modern optoelectronics, and is ubiquitous in a wide range of applications from data processing and telecommunications to energy conversion. Analogously, actively tailoring the flow of heat is of fundamental importance in thermal management and recycling. Significant volume of work has focused on radiative heat transfer, where active modulation relies on controlling photon heat flux \cite{CarminatiGreffet_PRL2000, Laroche_NFHT_TPV, Mulet_SiC2002, FAN_review,Wei_Radiative_Review,Greffet_Nature,Padilla_PRL_Blackbody,JJ_PNAS,DeSoyza_NatPhot}. In contrast to the far field, where the Stefan-Boltzmann law of thermal radiation imposes an upper bound to the amount of photon heat flux exchanged between two objects, in the near-field, evanescent waves tunneling can yield significantly larger heat exchange \cite{CarminatiGreffet_PRL2000, Raschke_NFHTReview, Basu_NFHTtreatment, Joulain_NFHT, Lipson_NFHT}. Thus, it is of importance to control radiative heat flux in the near-field, with potential applications in thermal circuitry \cite{Abdallah_ThermalTransistor, Li_ThermalTransistor}, thermal energy harvesting and recycling \cite{Zhao_NFTPV, Joannopoulos_SqueezingNFHT, Laroche_NFHT_TPV, Park_TPV2008}, refrigeration \cite{Rodriguez_Refrigeration2018} and nanoscale thermal imaging \cite{Kittel_ThermalImaging, DeWilde_Nature_NFHTImaging}.}

\par{Since the 1950's, the metal-oxide-semiconductor (MOS) device has lied at the heart of solid-state electronic operations \cite{Shockley_transistor, Engelbrecht_SiMOS}. In a MOS device, such as the metal-oxide-semiconductor field effect transistor (MOSFET), voltage gating of a semiconductor against a metallic electrode separated by an oxide layer results in accumulation and depletion of charge carriers. This in turn leads to changes in the electrical conductivity of the active region and hence the operation of an electrical switch.}

\par{Analogously, to create a switch for photon-based heat flux, one will need materials for which their photonic properties can be tuned. Previous considerations in radiative thermal switching have focused on phase change materials, such as vanadium oxide, or phase change ferroelectric materials such as perovskites, as crystallographic phase changes are typically accompanied by changes in optical properties \cite{Abdallah_ThermalTransistor, Li_ThermalTransistor, Zwol_PhaseChange2011,Basov_VO2Science2007,Zwol_PhaseChangeVO22011,Dyakov_NFHTVO2,VO2NFHT_Novotny2016,Iizuka_VO2NFHT,Boriskina_Ferroelectric2015}. However, despite the pronounced crystallographic changes occurring near a phase transition, changes in optical properties, e.g. dielectric permittivity, are typically moderate, leading to relatively modest contrast ratios, for example a $17 \%$ tunability range in photon heat flux reported in \cite{Boriskina_Ferroelectric2015}. Alternative mechanisms for tuning near-field heat transfer include photo-excitation \cite{Minnich_OpticalPumpingNFHT} and non-linear processes \cite{Khandekar_FourwavemixingAPL}, that, however, require large optical power densities beyond the $\mathrm{kW}/\mathrm{cm}^2$ range. Graphene's carrier density-tunable optical response has also been proposed as a heat flux modulating mechanism \cite{Ilic_Graphene2012,Ilic_grapheneACS, Zhao_graphenePRB,Lim_Graphene2013,Hong_SiChBNgraphene2018,Messina_GraphenePRB2017}, however a realistic configuration for practical carrier injection, and corresponding electrostatic considerations, have not been discussed.}

\begin{figure}[]
\centering
 \includegraphics[width=1\linewidth]{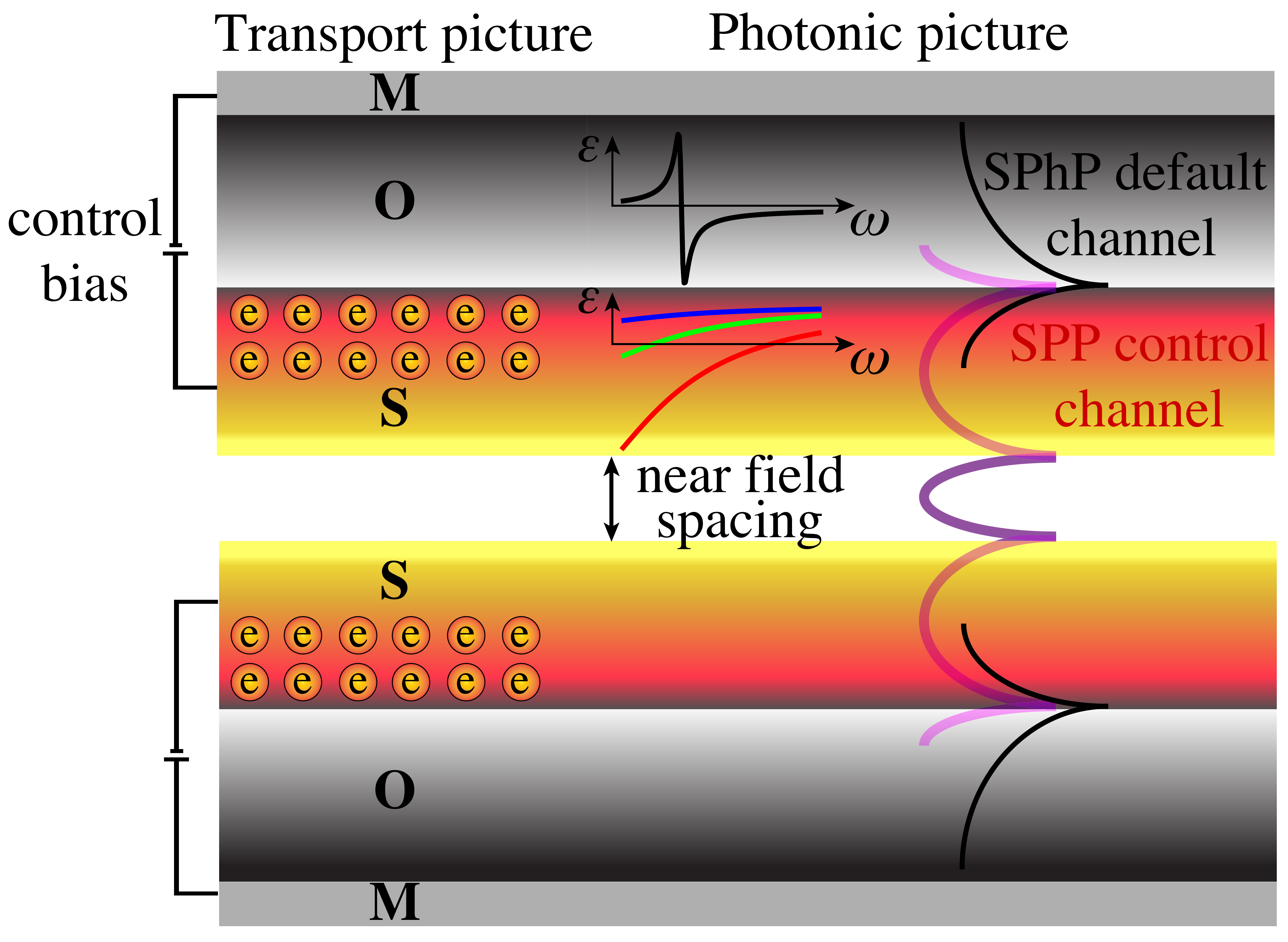}
  \caption{\textbf{Thermal metal (M)-oxide (O)-semiconductor (S) switch.} Application of an external voltage across the oxide can alter the charge carrier density in the vicinity of the semiconductor/oxide interface. Schematic depicts either n-doped semiconductor in accumulation mode or p-doped semiconductor in depletion mode. Most oxides exhibit optical phonons in their crystal structure, that yield Lorentz-type polaritonic permittivity resonances (black $\epsilon$ curve), and corresponding surface phonon polariton (SPhP) evanescent modes. Free charge carriers in the semiconductor/oxide interface yield a metallic response that depends on the carrier density. Green permittivity curve corresponds to absence of control bias, whereas blue and red curves correspond to operation in depletion and accumulation, respectively. A tunable surface plasmon polariton (SPP) channel in the active region controls the thermal properties of the module.}
 \label{fig:Figure1}
\end{figure}
  
\par{In this paper, we show that the concept of an MOS capacitor, as is widely used in electronic switching, can be alternatively used as a thermal MOS switch that achieves significant switching of photon heat flux by controlling a gate bias. In terms of contrast ratios, our thermal MOS switch significantly outperforms previous proposals, with switching (ON/OFF) ratios reaching $225\%$ in some of our designs. Analogously to the electronic MOS, the underlying mechanism lies in controlling the charge carrier density in an ultra-thin heavily doped semiconductor layer gated against a metallic electrode. Changes in carrier density in such an ultra-thin layer affect the plasmon supported by the layer (Fig. \ref{fig:Figure1}). By placing two such MOS capacitors in close proximity, as shown in Fig. \ref{fig:Figure1}, the heat transfer between the two capacitors is then actively controlled via a gate bias. In addition to significant contrast ratio, our proposed scheme can lead to CMOS compatible near-field thermal switching modules, with switching speed that coincides with that of an electronic switch at the GHz level \cite{Speed_fieldeffect1, Speed_fieldeffect2}. Additionally, operation of the proposed thermal MOS switch requires very low electrostatic power densities of the order of tens of $\mathrm{W}/\mathrm{cm}^2$ for typical MOS device areas that can reach hundreds of $\mu m^2$ \cite{Papadakis_FieldEffectPRB, MOS_devicesize_2003}.}

\par{Although we refer to the proposed scheme as a thermal ``MOS'' switch, we note that the ``semiconductor'' may be any plasmonic material with tunable carrier density. We explore three classes of gate-tunable materials: transparent conductive oxides (TCOs) and nitrides, doped semiconductors, and monolayer graphene, which support surface plasmon polaritons (SPPs) below their plasma frequency. The ``oxide'' layer may be any insulating material, and may also support a phonon-polaritonic dielectric response. The oxide layer thus may also support a surface phonon polariton (SPhP), which plays a significant role in the near-field heat transfer as well. In this work, we use SiC as the ``oxide'' layer. And moreover, we also consider near-field heat transfer between two SiC regions as the benchmark with which we compare our thermal switch. The reason for the selection of SiC is two-fold: (i) SiC exhibits a high-quality factor resonance \cite{Cadwell_SiC} that aligns with the peak of blackbody radiation at moderate temperatures, and its SPhP has been widely exploited in the study of near-field heat transfer \cite{Fan_SiC, Joulain_SiC, SiC_PRL1959, Boriskina_NFHT2015, Basu_NFHTtreatment, CarminatiGreffet_PRL2000, Raschke_NFHTReview, Joulain_NFHT, Mulet_SiC2002, Lipson_NFHT, Wang_NFHTSiCfilms}. (ii) Apart from its optimal thermal response, SiC is also a good electrical insulator with a breakdown voltage of $E_\mathrm{br}=3$ MV/cm (for 4H-SiC, while $E_\mathrm{br}=3.2$ MV/cm for 6H-SiC along its optical axis) \cite{Qin_SiCbreakdown, Pensl_SiCbreakdown, Chelnokov_SiCbreakdown1997, NeudeckSiCbreakdown}, hence it can sustain large electrostatic biases in serving as the ``oxide'' in the thermal MOS configuration of Fig. \ref{fig:Figure1}. Aside from SiC, our results can be generalized to other choices of oxides such as silicon dioxide (SiO\textsubscript{2}), or hexagonal boron nitride (hBN) \cite{Zhao_graphenePRB, Hong_SiChBNgraphene2018}.}

\section{\label{Formulation}Formulation}

\par{To treat the heat transfer in the geometry shown in Fig. \ref{fig:Figure1}, we employ the standard fluctuational electrodynamics approach. The total heat flux per area exchanged between two bodies at temperatures $T_\mathrm{1}>T_\mathrm{2}$, separated by a vacuum gap of thickness $d$, is given by \cite{CarminatiGreffet_PRL2000, Basu_NFHTtreatment, Laroche_NFHT_TPV, Raschke_NFHTReview}}
\begin{equation}\label{eq:1}
Q=\frac{1}{4\pi^2}\int d\omega[\Theta(\omega,T_\mathrm{1})-\Theta(\omega, T_\mathrm{2})] S(\omega)
\end{equation}
{where $\Theta(\omega,T)=\hbar \omega/[\mathrm{exp}(\hbar \omega/k_\mathrm{B}T)-1]$ is the mean energy of a harmonic oscillator at frequency $\omega$ and temperature $T$, and $S(\omega)$ is the spectral transfer function. Through the paper, unless otherwise noted, we set $T_\mathrm{1}=600$ K, and $T_\mathrm{2}=300$ K. For planar configurations,}
\begin{equation}\label{eq:2}
S(\omega)=\int d\beta    \xi(\omega,\beta)\beta
\end{equation}
{where $\beta$ is the in-plane wavenumber. The parameter $\xi(\omega,\beta)$ is the photon transmission probability averaged over transverse electric (s) and transverse magnetic (p) polarizations. For separation distances $d$ smaller than the thermal wavelength $\lambda_\mathrm{T}=b/T$, where $b=2989$ $\mu$m$\cdot$K, the evanescent wave's contribution dominates, and}
\begin{equation}\label{eq:3}
\xi(\beta>k_\mathrm{o},\omega)=\frac{4\mathrm{Im}(r_\mathrm{1})\mathrm{Im}(r_\mathrm{2})e^{-2|k_\mathrm{z0}|d}}{|1-r_\mathrm{1}r_\mathrm{2}e^{-2i k_\mathrm{z0}d}|^2}
\end{equation}
{In Eq. (\ref{eq:3}), $k_\mathrm{o}=\omega/c$ is the free space wavenumber, $r_\mathrm{1,2}$ are the Fresnel reflection coefficients for incidence from the gap (vacuum) to bodies 1 and 2, and $k_\mathrm{z0}$ is the out-of-plane wavenumber in vacuum.}

\section{\label{BasicConcept}Concept}

\par
\begin{figure}[]
\centering
 \includegraphics[width=1\linewidth]{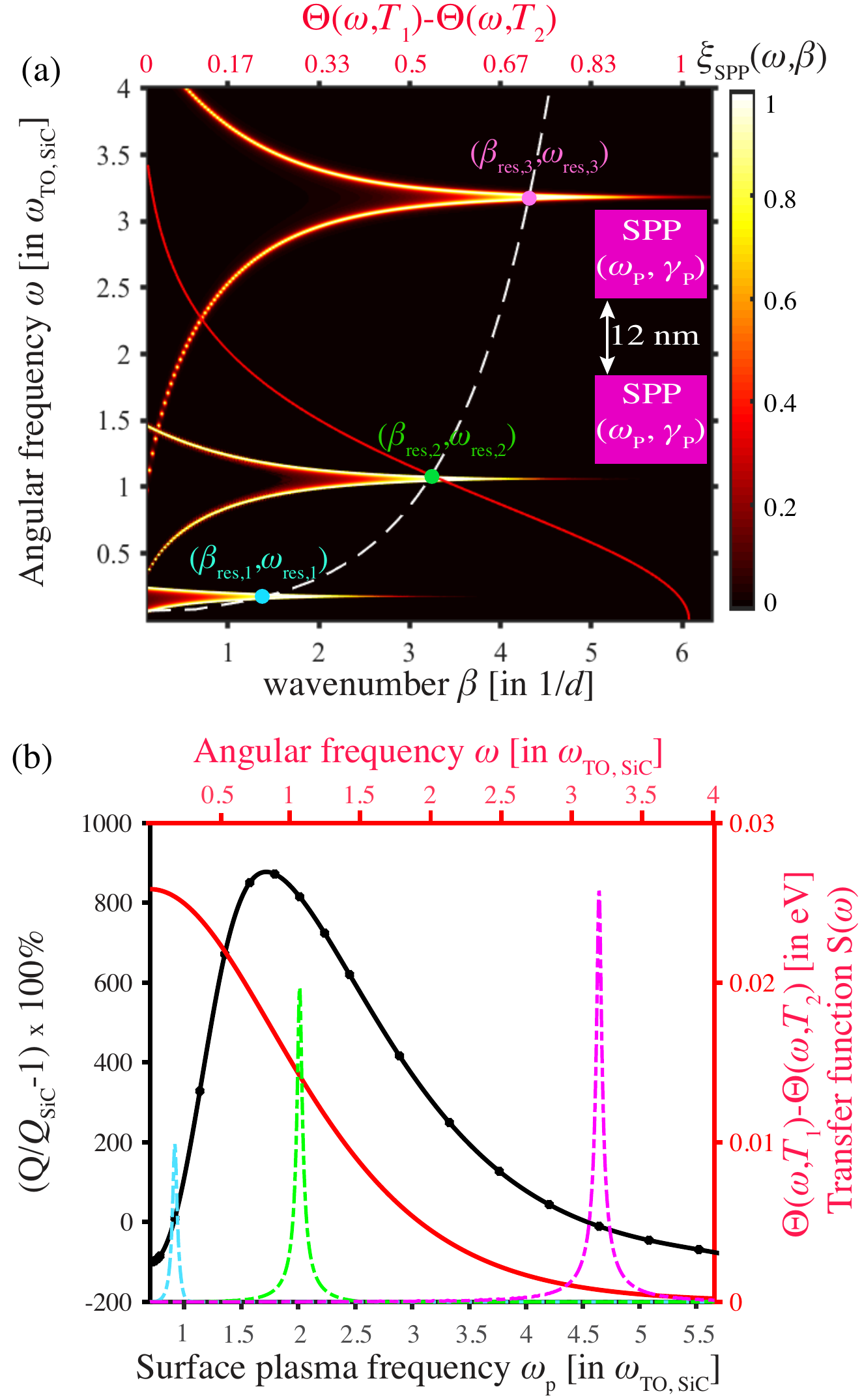}
  \caption{\textbf{Near-field heat transfer between two bulk plasmonic regions as described by the Drude model.} (a) Photon transmission probability $\xi_\mathrm{SPP}$ for a pair of semi-infinite Drude metals separated by $d=12$ nm, with $\omega_\mathrm{P1}=0.25\omega_\mathrm{TO,SiC}$ (lower curve), $\omega_\mathrm{P2}=1.5\omega_\mathrm{TO,SiC}$ (middle curve) and $\omega_\mathrm{P3}=4.5\omega_\mathrm{TO,SiC}$ (upper curve), for $\gamma_\mathrm{p}=5\gamma_\mathrm{SiC}$ . Red curve (upper horizontal axis) corresponds to the energy exchange per photon $\Theta(\omega,T_\mathrm{1})-\Theta(\omega,T_\mathrm{2})$, for $T_\mathrm{1}=600$ K, $T_\mathrm{2}=300$ K, normalized to its value at $\omega=0$. White dashed line corresponds to Eq. \ref{eq:6}, the wavenumber $\beta_\mathrm{res}$ that maximizes near-field heat transfer, as a function of $\omega_\mathrm{p}$ (the considered range of $\omega_\mathrm{p}$ is the same as that of $\omega$). Coordinates ($\beta_\mathrm{res,i}$, $\omega_\mathrm{res,i}$) for $i=1,2,3$ correspond to the wavenumber and frequency that maximize $\xi_\mathrm{SPP}$ (Eqs. \ref{eq:5}, \ref{eq:6}), for $\omega_\mathrm{P1}$, $\omega_\mathrm{P2}$, $\omega_\mathrm{P3}$. (b) Black curve (left vertical axis and corresponding lower horizontal axis): total heat flux (relative to SiC) for increasing $\omega_\mathrm{p}$. Red curve (right vertical axis and corresponding upper horizontal axis): $\Theta(\omega,T_\mathrm{1})-\Theta(\omega,T_\mathrm{2})$ in eV. Dashed curves (right vertical axis and corresponding upper horizontal axis): $S(\omega)$ (Eq. \ref{eq:2}), calculated for $\omega_\mathrm{P1}$ (cyan), $\omega_\mathrm{P2}$ (green) and $\omega_\mathrm{P3}$ (magenta). The peak value for the magenta curve is $1.03\times10^{16}$ $\mathrm{m}^{-2}$ .}
 \label{fig:Figure2}
\end{figure}

{In this section, we demonstrate that the total heat flux $Q$ exchanged between a pair of semi-infinite plasmonic materials separated by a vacuum gap in the near-field, as shown in the inset of Fig. \ref{fig:Figure2}a, depends on the plasma frequency, and therefore depends on the carrier density. This result enables the use of carrier injection to tune near-field heat transfer.}

\par{We model the plasmonic dielectric function with the Drude model}
\begin{equation}\label{eq:Drude}
\epsilon(\omega)=\epsilon_\mathrm{\infty}-\frac{\omega^2_{\mathrm{p}}}{\omega^2+i\omega\gamma_\mathrm{p}}
\end{equation}
where $\omega_\mathrm{p}$ is the plasma frequency, $\gamma_\mathrm{p}$ is the momentum-relaxation rate of charge carriers, and $\epsilon_\mathrm{\infty}$ is the high-frequency dielectric constant. The plasma frequency is given by $\omega_\mathrm{p}=\sqrt{N e^2/\epsilon_\mathrm{o} m^{*}}$, where $e$ is the electron charge, $\epsilon_\mathrm{o}$ is the free space dielectric constant, $m^{*}$ is the effective mass of charge carriers and $N$ is the charge carrier density. Below $\omega_\mathrm{p}$, the plasmonic interfaces support p-polarized SPPs, while the contribution from s-polarization may be ignored as there is no s-polarized surface wave in this system \cite{Raschke_NFHTReview, Mulet_SiC2002, Abdallah_betaSPP, Papadakis_Magnetic}. Near the SPP resonance, where the in-plane wavenumber $\beta$ is large, the Fresnel coefficient becomes $r_\mathrm{1,2}\approx \frac{\epsilon(\omega)-1}{\epsilon(\omega)+1}$, and the photon transmission probability of Eq. \ref{eq:3} may be approximated by \cite{Abdallah_betaSPP}
\begin{gather*}\label{eq:4}
\xi_\mathrm{SPP}(\omega)=\frac{16\omega_\mathrm{p}^4\gamma_\mathrm{p}^2\omega^2}{\Pi(\omega, x)} \text{ where }\\
\Pi(\omega, x)=16[(\omega_\mathrm{p}^{2}/2-\omega^2)^2+\gamma_\mathrm{p}^2\omega^2]^2e^{2x}- \\ 
8[(\omega_\mathrm{p}^{2}/2-\omega^2)^2-\gamma_\mathrm{p}^2\omega^2]\omega_\mathrm{p}^4+\omega_\mathrm{p}^8e^{-2x}
\end{gather*}
{where we have set $\epsilon_\mathrm{\infty}=1$, and $x=\sqrt{\beta^2-k_\mathrm{o}^2}d$. When $d$ is smaller than the thermal wavelength, near-field heat transfer is dominated by contributions from evanescent waves with $\beta\gg k_\mathrm{o}$, hence $x\approx \beta d$. The photon transmission probability $\xi_\mathrm{SPP}$ is upper bounded by unity. By setting $\xi_\mathrm{SPP}$ to unity, with some algebraic manipulation (see \cite{Abdallah_betaSPP} for details), we obtain the corresponding wavenumber}
\begin{equation}\label{eq:4}
\beta_\mathrm{SPP}(\omega)=\frac{1}{2d}\mathrm{ln}[\frac{\omega_\mathrm{p}^4/4}{(\omega_\mathrm{p}^2/2-\omega^2)^2+\gamma_\mathrm{p}^2\omega^2}]
\end{equation}
\par{In the integrand of Eq. \ref{eq:2}, which determines the spectral transfer function, due to the factor $\beta$, the maximum wavenumber where the photon transmission probability remains at unity strongly influences the total magnitude of heat transfer. To determine such a maximum wavenumber we set $\partial\beta_\mathrm{SPP}/\partial\omega=0$ to obtain}
\begin{equation}\label{eq:5}
\omega_\mathrm{res}=\sqrt{\frac{\omega_\mathrm{p}^2-\gamma_\mathrm{p}^2}{2}}
\end{equation}
{and, correspondingly, from Eq. \ref{eq:4},}
\begin{equation}\label{eq:6}
\beta_\mathrm{SPP}(\omega_\mathrm{res})=\beta_\mathrm{res}=\frac{1}{d}\mathrm{ln}[\big(\frac{\omega_\mathrm{p}}{\gamma_\mathrm{p}}\big)\frac{1}{2-(\gamma_\mathrm{p}/\omega_\mathrm{p})^2}]
\end{equation}
{Here, we refer to the frequency and wavenumber thus obtained (Eqs. \ref{eq:5}, \ref{eq:6}) as the resonant frequency and wavenumber, since, as we will see below, the spectral heat flux strongly peaks at such frequencies.}

\par{Eq. \ref{eq:6} demonstrates that, for $\omega_\mathrm{p}$ considerably larger than $\gamma_\mathrm{p}$, the dependence of $\beta_\mathrm{res}$ on $\omega_\mathrm{p}$ is logarithmic. This trend is shown in Fig. \ref{fig:Figure2}a with the photon transmission probabilities for a pair of semi-infinite plasmonic materials separated by a 12 nm vacuum gap. As mentioned in the introduction, we use SiC parameters as points of reference henceforth. Namely, we normalize the angular frequency, $\omega$, to $\omega_\mathrm{TO, SiC}=1.49\times10^{14}$ rad/s, the transverse optical phonon frequency of SiC \cite{Fan_SiC, Joulain_SiC, SiC_PRL1959}, and the total heat flux, $Q$, to the equivalent configuration separating two semi-infinite SiC plates, $Q_\mathrm{SiC}$. In Fig. \ref{fig:Figure2}a, we consider $\gamma_\mathrm{p}=5\gamma_\mathrm{SiC}$, where $\gamma_\mathrm{SiC}=8.97\times10^{11}$ rad/s is the phonon relaxation rate in SiC, and consider various plasma frequencies. For each plasma frequency, we plot the photon transmission probability $\xi(\omega,\beta)$. The position of $\omega$-$\beta$ pairs where $\xi(\omega,\beta)$ reaches unity thus appears as a contour in the $\omega$-$\beta$ plane. In Fig. \ref{fig:Figure2}a, the plasma frequency is set to $\omega_\mathrm{P1}=0.25\omega_\mathrm{TO, SiC}\approx 8\gamma_\mathrm{p}$ (lower curve), $\omega_\mathrm{P2}=1.5\omega_\mathrm{TO, SiC}\approx 50\gamma_\mathrm{p}$ (middle curve), and $\omega_\mathrm{P3}=4.5\omega_\mathrm{TO, SiC}\approx 150\gamma_\mathrm{p}$ (top curve). On each contour, the $\omega$-$\beta$ coordinates where the largest $\beta$ occurs agree well with Eq. \ref{eq:6}.}

\par{The analysis above on $\omega_\mathrm{res}$, in combination with the form of Eq. \ref{eq:1}, points to the mechanisms through which changing of $\omega_\mathrm{p}$ can influence the heat transfer. The spectral transfer function $S(\omega)$, as shown by the dashed lines in Fig. \ref{fig:Figure2}b, for the same $\omega_\mathrm{p}$'s as the ones considered panel (a), peaks strongly near $\omega_\mathrm{res}$. Moreover, the peak value of $S(\omega)$ increases as a function of $\omega_\mathrm{p}$, due to the increase of $\beta_\mathrm{res}$ as discussed above. On the other hand, the energy exchange per photon, $\Theta(\omega,T_\mathrm{1}) - \Theta(\omega,T_\mathrm{2})$, maximizes at $\omega =0$, and gradually decreases as $\omega$ increases (red curve in Fig. \ref{fig:Figure2}b). Consequently, the near-field heat flux $Q$ between the two bodies increases with $\omega_\mathrm{p}$ in the regime of small $\omega_\mathrm{p}$, and decreases with $\omega_\mathrm{p}$ in the regime of large $\omega_\mathrm{p}$, as shown by the black curve in Fig. \ref{fig:Figure2}b. Since $\omega_\mathrm{p}$ is directly related to the carrier concentration, the results here thus show that it is possible to drastically influence the near-field heat transfer in a plasmonic system by modulating the carrier density.} 

\begin{figure*}[]
\centering
 \includegraphics[width=1\linewidth]{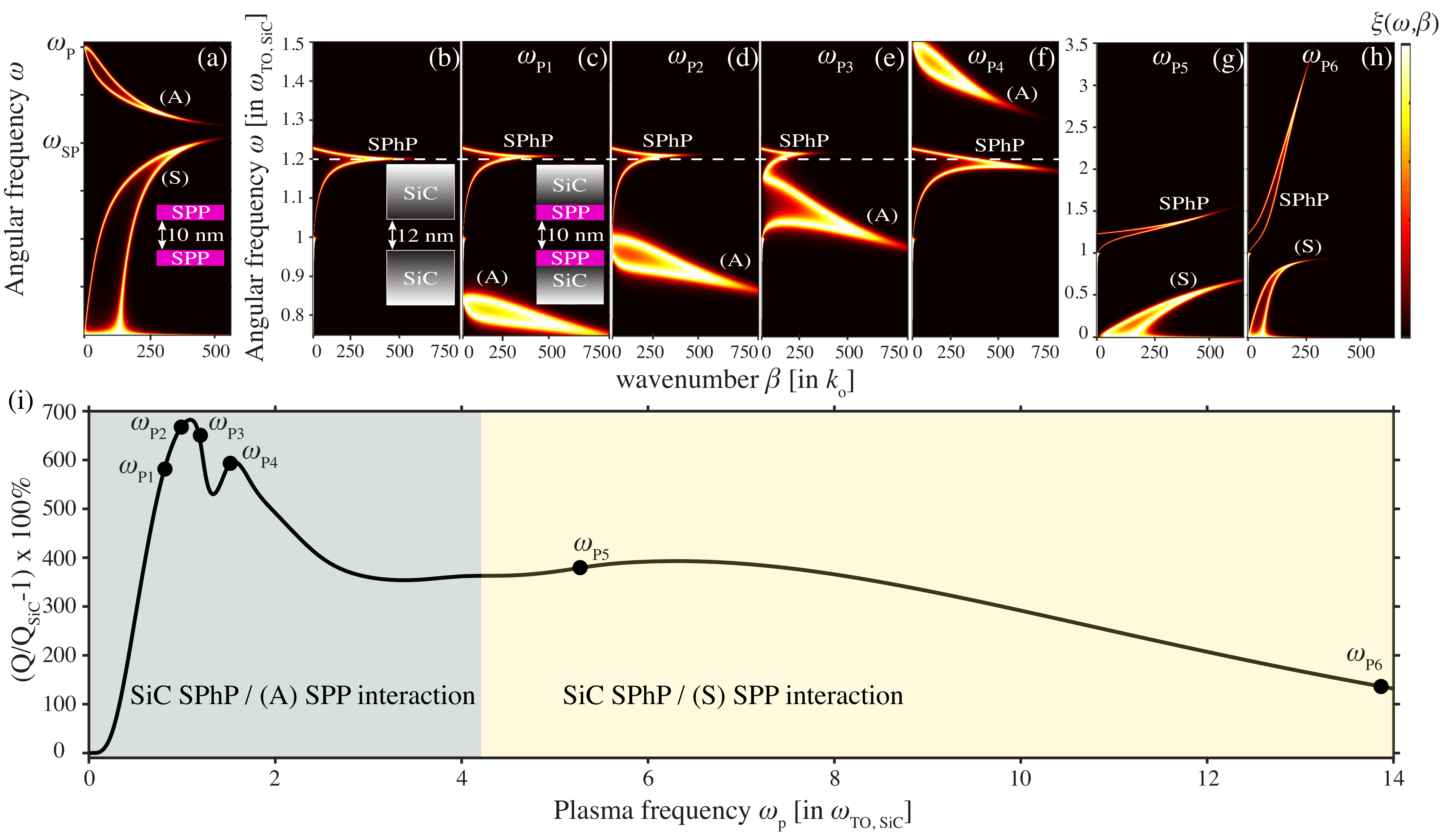}
  \caption{\textbf{Near-field heat transfer between plasmonic thin films on SiC.} (a) Photon transmission probability $\xi$ for a pair of ultra-thin plasmonic films separated by 10 nm, as depicted in the inset. Letters (A) and (S) refer to anti-symmetric and symmetric electric field distribution with respect to the center of each plasmonic film. (b) Photon transmission probability for a pair of semi-infinite SiC slabs separated by a 12 nm vacuum gap, as depicted in the inset. (c)-(h) Photon transmission probabilities for a pair of plasmonic thin films of thickness 1 nm on SiC, separated by  a 10 nm vacuum gap, as depicted in the inset of (c). The momentum-relaxation rate is set at $\gamma_\mathrm{p}=5\gamma_\mathrm{SiC}$, while the plasma frequency is $\omega_\mathrm{P1}=0.85\omega_\mathrm{TO, SiC}$ (c), $\omega_\mathrm{P2}=\omega_\mathrm{TO, SiC}$ (d), $\omega_\mathrm{P3}=1.15\omega_\mathrm{TO, SiC}$ (e), $\omega_\mathrm{P4}=1.5\omega_\mathrm{TO, SiC}$ (f), $\omega_\mathrm{P5}=5.5\omega_\mathrm{TO, SiC}$ (g), $\omega_\mathrm{P6}=14\omega_\mathrm{TO, SiC}$ (h). (i) Total heat flux (relative to SiC) for the heterostructure in the inset of panel (c), for increasing $\omega_\mathrm{p}$ of the plasmonic film, and for $T_\mathrm{1}=600$ K, $T_\mathrm{2}=300$ K. The grey-highlighted regime corresponds to the frequency range where the SiC SPhP interacts with the (A) branches of the SPP, and the relevant photon transmission probabilities  are displayed in panels (c)-(f), whereas the yellow-highlighted regime corresponds to the interaction of the SiC SPhP with the (S) SPP branches, and the relevant transmission probabilities are in panels (g)-(h).}
 \label{fig:Figure3}
\end{figure*}

\section{\label{Thin_Film}Interaction of surface plasmon and surface phonon polariton}

\par{In the previous section, we have shown that in a plasmonic system, changing the plasma frequency can lead to substantial modulation and switching of near-field heat transfer. In practice, the thermal MOS switch, as shown in Fig. \ref{fig:Figure1}, represents the most practical way to tune the plasma frequency of a heavily doped semiconductor layer. Compared with the idealized geometry as shown in the previous section, however, the thermal MOS switch in Fig. \ref{fig:Figure1} differs in two significant aspects: (i) The heavily doped region is of necessity a thin film, so that a modest voltage applied can lead to substantial change in the plasma frequency. (ii) There is a need for the oxide region for voltage gating purposes. As we will discuss in this section, both of these aspects significantly affect the characteristics of near-field heat transfer in the MOS configuration as compared to that between two bulk plasmonic regions.}

\par{We consider near-field heat transfer between a pair of plasmonic thin film first (inset in Fig. \ref{fig:Figure3}a). The peaks of the photon transmission probability consist of two main branches, labeled with (S) and (A) in Fig. \ref{fig:Figure3}a, each of which consists of two secondary branches. The labels (S) and (A) pertain to an electric field distribution that is symmetric and anti-symmetric, respectively, with respect to the center of the plasmonic film, whereas within each main branch the lower (upper) frequency secondary branch has a symmetric (anti-symmetric) field distribution with respect to the center of the gap.}

\par{In Fig. \ref{fig:Figure3}b, we consider a pair of semi-infinite SiC interfaces separated by a 12 nm vacuum gap. We model the permittivity of SiC with the Lorentz model \cite{Fan_SiC, Joulain_SiC, SiC_PRL1959}, $\epsilon_{\mathrm{SiC}}(\omega)=$\\
$\epsilon_{\mathrm{\infty}}[({\omega^2_{\mathrm{LO}}-\omega^2-i\gamma\omega})/({\omega^2_{\mathrm{TO}}-\omega^2-i\omega\gamma})]$, where $\epsilon_{\mathrm{\infty, SiC}}=6.7$, $\omega_\mathrm{LO, SiC}=1.83\times10^{14}$ rad/s, and $\gamma_\mathrm{SiC}=8.97\times10^{11}$ rad/s, $\omega_\mathrm{TO, SiC}=1.49\times10^{14}$ rad/s, as mentioned above. The transmission probability $\xi$ reaches unity along the SPhP modes of the coupled SiC-vacuum interfaces. The frequency range $[1,1.2]\omega_\mathrm{TO, SiC}$ of these modes is the SiC Reststrahlen band. $\beta_\mathrm{res}$ occurs near the frequency $\omega\sim1.2\omega_\mathrm{TO,SiC}$, where $\epsilon_\mathrm{SiC}\approx1$.}

\par{The characteristics of near-field heat transfer in the structure of Fig. \ref{fig:Figure1} can be understood in terms of the interaction between the bulk SiC SPhP modes with the plasmonic modes of the thin film. We start from the bulk SiC configuration and add two plasmonic films of thickness 1 nm atop of SiC, so that the vacuum gap is now set at 10 nm as shown in the inset of Fig. \ref{fig:Figure3}c. The parameter $\omega_\mathrm{p}$ of the plasmonic film is set to $\omega_\mathrm{P1}=0.85\omega_\mathrm{TO,SiC}$, while $\gamma_\mathrm{p}$ remains at $5\gamma_\mathrm{SiC}$ as in Figs. \ref{fig:Figure2}, \ref{fig:Figure3}a. Due to the low $\omega_\mathrm{p}$ in Fig. \ref{fig:Figure3}c, only the (A) SPP branch is visible, and the (S) branch lies at lower frequencies. The SPhP mode at the Reststrahlen band of SiC is not affected by the (A) SPP mode. By increasing the plasma frequency to $\omega_\mathrm{P2}=\omega_\mathrm{TO, SiC}$, the (A) SPP mode blueshifts (Fig. \ref{fig:Figure3}d) until, at $\omega_\mathrm{P3}=1.15\omega_\mathrm{TO, SiC}$, the SiC SPhP mode overlaps with the (A) SPP (Fig. \ref{fig:Figure3}e). We note that even though the bands overlap, the shapes of the SPhP modes and the SPP modes remain largely unchanged, indicating the weak interaction between the (A) SPP branch and the SPhP modes. Increasing the plasma frequency further, to $\omega_\mathrm{P4}=1.5\omega_\mathrm{TO, SiC}$, blueshifts the (A) SPP mode to frequencies beyond the Reststrahlen band of SiC (Fig. \ref{fig:Figure3}f).}

\par{Fig. \ref{fig:Figure3}i displays the enhancement in total integrated near-field heat transfer, $Q$, of the plasmonic film on SiC (inset in Fig. \ref{fig:Figure3}c) with respect to bulk SiC (inset in Fig. \ref{fig:Figure3}b), as a function of $\omega_\mathrm{p}$. The regime of $\omega_\mathrm{p}<4\omega_\mathrm{TO,SiC}$, highlighted with grey, corresponds to the plots discussed above (panels (c)-(f)), for which the SiC SPhP overlaps with the (A) SPP branch. It can be seen that by placing an ultra-thin plasmonic film on SiC one can enhance the net near-field heat transfer by almost $700\%$, by appropriately positioning the plasma frequency $\omega_\mathrm{p}$ with respect to $\omega_\mathrm{TO, SiC}$. This enhancement is the result of an additional channel for transferring heat, namely the (A) SPP branch, aside from the fundamental SPhP channel of SiC. The dip in $Q$ at $\omega_\mathrm{p}\approx1.3\omega_\mathrm{TO,SiC}$ identifies the point at which the (A) SPP mode crosses over the Reststrahlen band of SiC \cite{Perez_2018arxiv} (Figs. \ref{fig:Figure3}e, f).}

\par{{In contrast to the near-field heat transfer, $Q(\omega_\mathrm{p})$, of the bulk case, which exhibits a single maximum as a function of $\omega_\mathrm{p}$ (Fig. \ref{fig:Figure2}b), a pair of thin plasmonic films on SiC supports an additional local maximum at high $\omega_\mathrm{p}>4\omega_\mathrm{TO}$, shown in the yellow highlighted regime in Fig. \ref{fig:Figure3}i. This second local maximum originates from the symmetric (S) SPP branch (see Fig. \ref{fig:Figure3}a) interacting with the SiC SPhP, and is more broadband in nature. In panels (g), (f) we display the photon transmission probability relevant to this regime, particularly for $\omega_\mathrm{P5}=5.5\omega_\mathrm{TO, SiC}$ and $\omega_\mathrm{P6}=14\omega_\mathrm{TO, SiC}$. In contrast to panels (c)-(f), there is significant interaction between the SiC SPhP with the (S) SPP branch. This response is previously seen in heterostructures involving polaritonic materials and graphene, as the monolayer nature of graphene eliminates the (A) SPP mode and only the (S) SPP interacts with SPhPs \cite{Zhao_graphenePRB, Zhao_grapheneJHT, Boltasseva_DrudeTCOs, Brongersma_ITO2013, Fu_DopedSiDrude,Engelbrecht_SiMOS}.} 

 \section{\label{Implementation}Practical Implementation}
 
\par{In Fig. \ref{fig:Figure3} we demonstrated that an ultra-thin plasmonic film on SiC may significantly enhance near-field heat transfer with respect to bulk SiC. Furthermore, we showed that changes in the plasma frequency $\omega_\mathrm{p}$ of the plasmonic film can tune near-field heat transfer considerably. Changes in $\omega_\mathrm{p}$ may be induced via modulation of charge carrier density. In search for substances with tunable carrier density, we resort to three classes of materials: TCOs and nitrides, doped semiconductors, and graphene.}

\par{In the class of TCOs we investigate ITO, which is traditionally used as a contact electrode in photovoltaic cells due to its simultaneous conducting and transparent nature. ITO is an emerging plasmonic material with considerably lower losses compared to noble metals, while also having gate-tunable carrier density \cite{Lee_Plasmonstor, Ghazaleh_ITO, Boltasseva_DrudeTCOs, Boltasseva_DrudeMetals,Boltasseva_OxidesNitrides1,Boltasseva_OxidesNitrides2,Brongersma_ITO2013}. In the class of semiconductors we study Si \cite{Fu_DopedSiDrude} as it is the most broadly used semiconductor in solid-state technology, ranging from solar harvesting to optoelectronics and silicon photonics. Graphene's monolayer thickness makes it an attractive candidate material for ultra-compact near-field heat transfer thermal switches. The carrier density of these materials may be actively modulated when electrostatically gated, as described in the general concept of Fig. \ref{fig:Figure1}, where they can be seen as the ``S'' of the thermal MOS switch. Below the active material lies a thick layer of SiC, which serves as the ``oxide'' of the thermal MOS switch. The SiC layer can sustain large gate voltages due to its large breakdown field of $E_\mathrm{br}=3$ MV/cm \cite{Qin_SiCbreakdown, Pensl_SiCbreakdown, Chelnokov_SiCbreakdown1997, NeudeckSiCbreakdown}, while also by itself contributing significantly to the heat transfer proparties of the system as we see in the previous section \cite{Fan_SiC, Joulain_SiC, SiC_PRL1959, Boriskina_NFHT2015, Basu_NFHTtreatment, CarminatiGreffet_PRL2000, Raschke_NFHTReview, Joulain_NFHT, Mulet_SiC2002, Lipson_NFHT, Wang_NFHTSiCfilms}. We set the thickness of the SiC layer to $\Delta=1$ $\mu$m for ensuring that (i) the SiC SPhP behaves like the bulk SPhP discussed in Fig. \ref{fig:Figure3}, and that (ii) large gate biases, $V_\mathrm{g}$, may be applied prior to electrical breakdown occurring. We consider silver as the back-side gating electrode (``M'' of the thermal MOS switch in Fig. \ref{fig:Figure1}).} 
  
\subsection{\label{TCOs}Transparent conductive oxides}

\par{We model ITO with the Drude model of Eq. \ref{eq:Drude} , using $m^{*}=0.35m_\mathrm{e}$, $\gamma_\mathrm{p}=1.4\times10^{14}$ rad/s, and $\epsilon_\mathrm{\infty}=3.9$ \cite{Lee_Plasmonstor, Papadakis_FieldEffectPRB, Brongersma_ITO2013,Zayats_PRLITO, Ghazaleh_ITO}. In Fig. \ref{fig:Figure4}a we display the active (red) and background regions (yellow) of ITO. When no bias is applied ($V_\mathrm{g}=0$), the active and background regions have the same carrier density $N_\mathrm{A}=N_\mathrm{B}$. An applied bias $V_\mathrm{g}$ causes accumulation ($N_\mathrm{A}>N_\mathrm{B}$) or depletion ($N_\mathrm{A}<N_\mathrm{B}$) of charge carriers in the active region, for which the plasma frequency $\omega_\mathrm{p}$ blueshifts and redshifts, respectively. The thickness of the active region is defined by the Thomas-Fermi screening theory \cite{Brongersma_ITO2013, Zayats_PRLITO}, and, to first order, can be approximated by the Debye length \cite{Nicollian_MOSbook,Papadakis_FieldEffectPRB}}
\begin{equation}\label{eq:7}
L_\mathrm{D}=\sqrt{\frac{\epsilon_\mathrm{o}\epsilon_\mathrm{o,SiC}k_\mathrm{B}T}{e N_\mathrm{A}}}
\end{equation}
\noindent{In Eq. \ref{eq:7} $\epsilon_\mathrm{o,SiC}$ is the DC dielectric constant of SiC set to 9.72 \cite{Choyke_SiC_DCproperties}, $k_\mathrm{B}$ is the Boltzmann constant, and $T$ is the temperature. The active region is typically ultra-thin, as shown for ITO at different carrier densities and temperatures in Fig. \ref{fig:Figure4}b, c.f. on the order of 1-5 nanometers \cite{Brongersma_ITO2013, Lee_Plasmonstor,Zayats_PRLITO,Brongersma_ITO2013}, compared to the wavelength of thermal radiation that lies in the regime of tens of microns. Despite the minuscule size of the Debye length, small changes to its optical properties ($\omega_\mathrm{p}$) via carrier density modulation can significantly impact near-field heat transfer. By minimizing the thickness of the background region in Fig. \ref{fig:Figure4}a, we maximize the range of tunability of near-field heat transfer by bringing the active region where changes of the optical properties take place as close to the vacuum gap as possible. SPPs and SPhPs that carry thermal energy across the vacuum gap (see Fig. \ref{fig:Figure1}) are strongly confined in the vicinity of the vacuum gap/active material interface and are, therefore, highly sensitive to changes of the optical properties of this region. Here, we set the total thickness of the ITO layer (background and active region in Fig. \ref{fig:Figure4}a) to 3 nm. The vacuum gap has thickness $d=10$ nm.}

\par{The maximum achievable carrier density in the active region is subject to the SiC breakdown constraint through}
\begin{equation}\label{eq:8}
\frac{V_\mathrm{g}}{\Delta}=\frac{q_\mathrm{A}}{\epsilon_\mathrm{o}\epsilon_\mathrm{o,SiC}} \leq E_\mathrm{br}
\end{equation}
\noindent{where $q_\mathrm{A}$ is the charge per area given by $q_\mathrm{A}=eN_\mathrm{A}L_\mathrm{D}$. From Eq. \ref{eq:8} we compute the maximum carrier density, beyond which SiC cannot sustain the applied electric field. The corresponding breakdown voltage is $V_\mathrm{g,br}=E_\mathrm{br}\Delta=300$ V.}

\begin{figure}[h!]
\centering
 \includegraphics[width=1\linewidth]{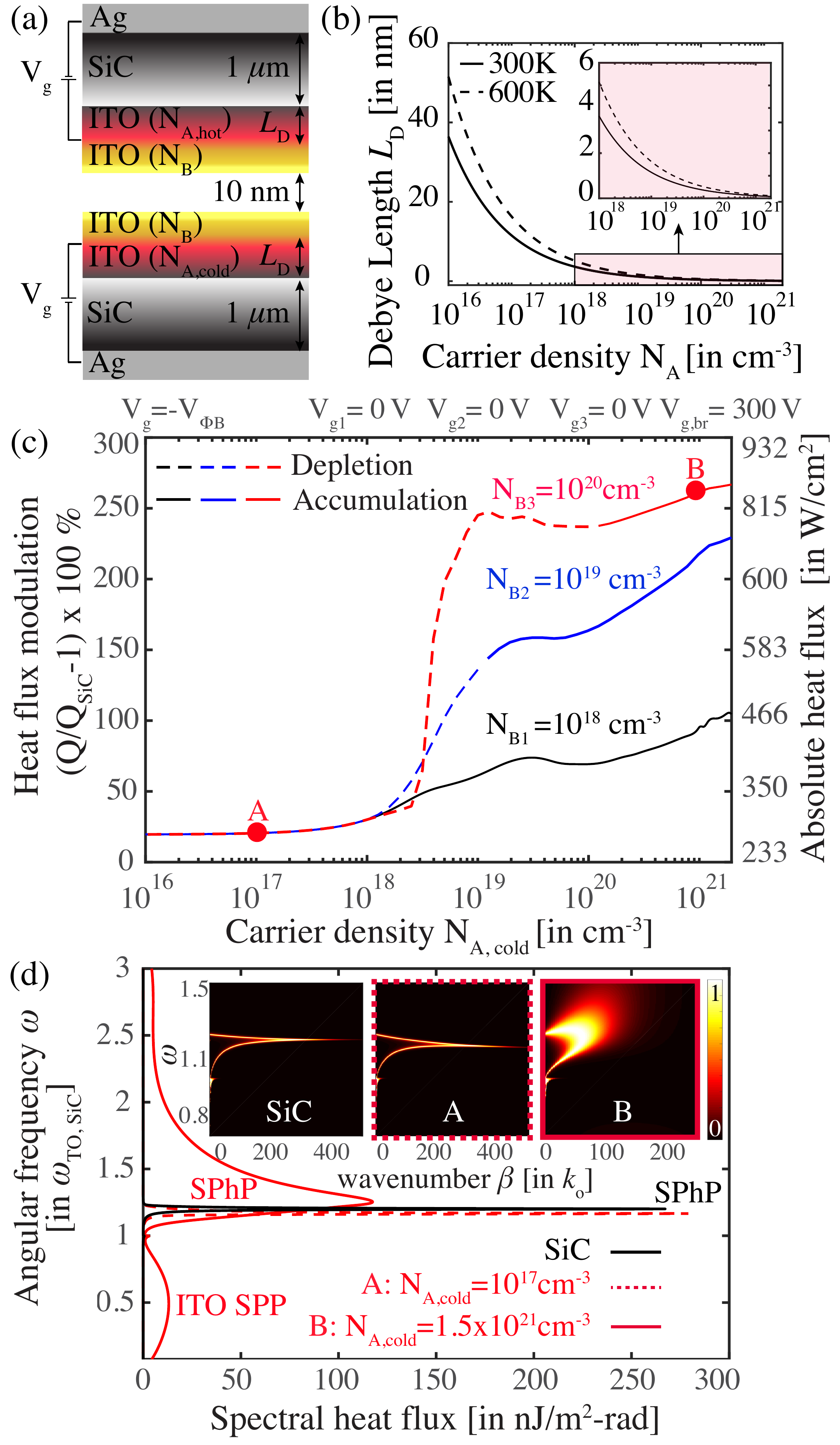}
  \caption{\textbf{Practical implementation of thermal MOS switch with ITO.} (a) Schematic of ITO-based thermal MOS switch. (b) Debye length for ITO at different temperatures. (c) Total heat transfer coefficient for $T_\mathrm{1}=600$ K, $T_\mathrm{2}=300$ K, relative to the SiC benchmark, with respect to the carrier density $N_\mathrm{A,cold}$ at the cold side. The background carrier density is $N_\mathrm{B}=10^{18}$, $10^{19}$, and $10^{20}$ $\mathrm{cm}^{-3}$ in results shown with black, blue, and red, respectively. Positive $V_\mathrm{g}$ corresponds to accumulation ($N_\mathrm{A}>N_\mathrm{B}$, solid lines), and negative $V_\mathrm{g}$ corresponds to depletion ($N_\mathrm{A}<N_\mathrm{B}$, dashed lines). The maximum considered carrier density  corresponds to the breakdown condition of SiC, $V_\mathrm{g,br}=300$ V. The blackbody limit lies at $(Q/Q_\mathrm{SiC}-1)\times 100 \%=-99.7$\%. (d) Spectral heat flux for carrier densities corresponding to points A and B in panel (c), in depletion and accumulation mode, respectively. Corresponding photon transmission probabilities are shown in the insets in depletion (middle) and accumulation (right). Results are compared to bulk SiC (black curve and left-most inset).}
 \label{fig:Figure4}
\end{figure}

\par{The minimum considered carrier density corresponds to the flat band condition $V_\mathrm{g}=-V_\mathrm{\Phi B}$, which sets the boundary between operation in inversion and depletion mode in an MOS \cite{Nicollian_MOSbook}. For $V_\mathrm{g}$ in the range $[-V_\mathrm{\Phi B},0]$, majority carriers, i.e. electrons for ITO, are depleted in the active region, and its optical response becomes more dielectric compared to the background. By contrast, for $V_\mathrm{g}$ in the range $[0,V_\mathrm{g,br}]$, majority carriers are accumulated in the active region, which becomes more metallic with respect to the background. Operation in inversion mode ($V_\mathrm{g}<-V_\mathrm{\Phi B}$) is not considered here, as it affects near-field heat transfer in the same way as in accumulation mode, with the role of majority carriers inverted.}

\par{The total calculated heat transfer, $Q$, as a function of the carrier density in the active region is shown in Fig. \ref{fig:Figure4}c. $Q$ is plotted normalized to the heat exchange between a pair of semi-infinite SiC surfaces separated by 16 nm, $Q_\mathrm{SiC}$, i.e. in the absence of the ITO thin films, whereas on the right vertical axis absolute values of power density are also displayed. The considered background carrier densities are $N_\mathrm{B}=10^{18},10^{19},10^{20}$ $\mathrm{cm}^{-3}$, shown with black, blue and red, respectively. Depletion mode is depicted with dashed lines, whereas accumulation mode is depicted with solid lines. As seen at $V_\mathrm{g}=V_\mathrm{g,br}$, near-field heat transfer is maximized for maximum carrier density in the background region, namely for $N_\mathrm{B}=10^{20}$ $\mathrm{cm}^{-3}$. This result is expected as explained in detail in Figs. \ref{fig:Figure2}, \ref{fig:Figure3}; increasing carrier density, and therefore increasing $\omega_\mathrm{p}$ leads to an increase in total near-field heat transfer, contributed by the SPP channel, as long as the mean energy exchange per photon overlaps with the SPP resonance frequency.}

\par{In Fig. \ref{fig:Figure4}c we demonstrated that adding an ultra-thin plasmonic film of ITO on SiC can induce significant enhancement to the neat heat transfer with respect to bulk SiC. Importantly, we showed that field effect modulation of the carrier density leads to active control of near-field heat transfer, which can vary with respect to SiC between 20 \% and 265 \%. By considering the OFF and ON states of this near-field heat transfer modulating scheme to correspond to $V_\mathrm{g}=0$ V and $V_\mathrm{g}=V_\mathrm{g,br}=300$ V, respectively, we obtain contrast ratios as high as 190 \%.} 

\par{In Fig. \ref{fig:Figure4}d we display the spectral heat flux, defined as the integrant quantity in Eq. \ref{eq:1}, with respect to bulk SiC, as discussed in panel (c). Results are shown for depletion (dashed lines) and accumulation (solid lines), corresponding to the points A and B in panel (c), respectively. The spectral heat flux for bulk SiC is shown with the black curve for comparison. Additionally, in the insets are the photon transmission probabilities ($\xi$) for bulk SiC (left) as well as for depletion (middle) and accumulation (right) modes of our system with ITO. In the depletion mode, the photon transmission probability and respective heat flux do not change considerably with respect to bulk SiC. This is explained by the low carrier density at point A in Fig. \ref{fig:Figure4}c that yields small $\omega_\mathrm{p}$ with respect to $\omega_\mathrm{TO, SiC}$, therefore SPP of ITO is too low in frequency to strongly interact with the Reststrahlen band of SiC or to contribute significantly to heat transfer. The slight increase in integrated heat flux at point A, with respect to bulk SiC, in Fig. \ref{fig:Figure4}c, results from the large plasmonic losses ($\gamma_\mathrm{p}$) in ITO, namely $\gamma_\mathrm{ITO}=[200-263]\gamma_\mathrm{SiC}$ \cite{Brongersma_ITO2013,Boltasseva_DrudeTCOs,Lee_Plasmonstor, Papadakis_FieldEffectPRB,Zayats_PRLITO, Ghazaleh_ITO}, for the range of carrier densities considered here. The ITO losses result in a slightly-broadened SPhP peak in spectral heat flux (dashed curve in panel (d)) with respect to bulk SiC (black curve in panel (d)), which, when integrated, results in a $Q$ larger than $Q_\mathrm{SiC}$.}

\par{In the accumulation mode, the ITO SPP branch blueshifts considerably, resulting in an additional broad peak in spectral heat flux, noted as ITO SPP in panel (d) of Fig. \ref{fig:Figure4} (solid red curve). The broadband nature of the ITO SPP peak is also the result of large plasmonic losses in ITO, which makes it hard to distinguish it from the SiC SPhP in the photon transmission probability on the right-side inset of Fig. \ref{fig:Figure4}d.}

\subsection{\label{Semiconductors}Doped semiconductors}

\begin{figure}[]
\centering
 \includegraphics[width=1\linewidth]{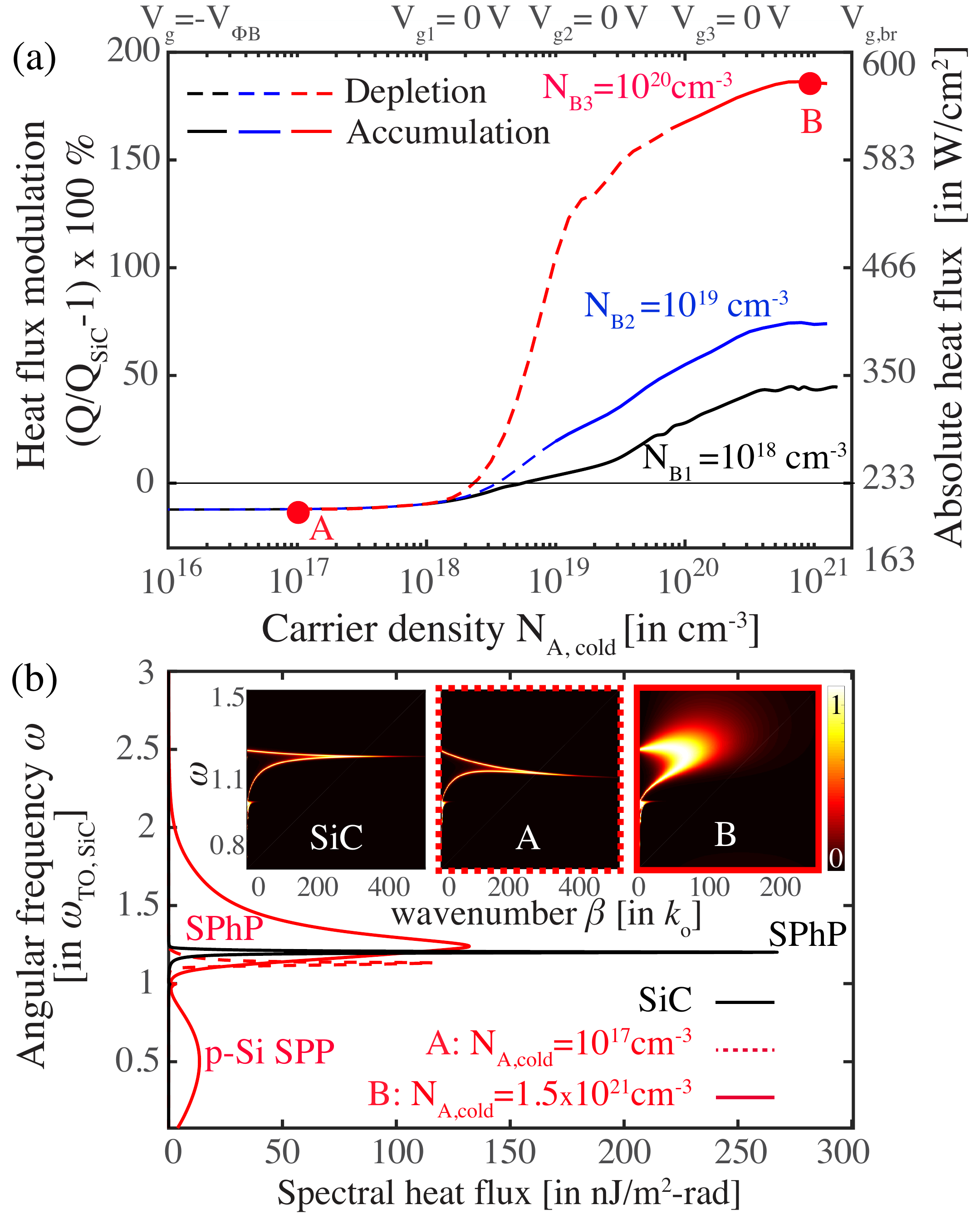}
  \caption{\textbf{Thermal MOS switch with p-doped Si.} (a) Total heat transfer coefficient for $T_\mathrm{1}=600$ K, $T_\mathrm{2}=300$ K, relative to the SiC benchmark, with respect to the carrier density $N_\mathrm{A,cold}$ at the cold side. The background carrier density is $N_\mathrm{B}=10^{18}$, $10^{19}$, and $10^{20}$ $\mathrm{cm}^{-3}$ in results shown with black, blue, and red, respectively. The blackbody limit lies at $(Q/Q_\mathrm{SiC}-1)\times 100 \%=-99.7$\%. (b) Spectral heat flux for carrier densities corresponding to points A and B in panel (a), in depletion and accumulation mode, respectively. Corresponding photon transmission probabilities are shown in the insets in depletion (middle curve) and accumulation (right curve). Results are compared to bulk SiC (black curve and left-most $\xi$ curve).}
 \label{fig:Figure5}
\end{figure}

\par{Next, we consider p-doped Si as the active material, in a configuration identical to that of Fig. \ref{fig:Figure4}a, where ITO is replaced by Si. We model doped Si with \cite{Fu_DopedSiDrude}, accounting for temperature effects on the dielectric function, namely on the effective mass of holes and electrons $m^{*}$, the momentum-relaxation rate $\gamma_\mathrm{p}$, and band gap. The total calculated heat transfer, $Q$, is shown in Fig. \ref{fig:Figure5}a, where dashed lines correspond to operation in the depletion mode, whereas solid lines correspond to the accumulation mode. In contrast to the system with ITO in Fig. \ref{fig:Figure4}, here, since Si is p-doped, accumulation and depletion correspond to increase and decrease in the number of holes as majority carriers in the active region, with respect to the background, respectively. With respect to bulk SiC, near-field heat transfer can vary between -12 \% and 185 \%. The ON ($V_\mathrm{g}=0$ V)/OFF ($V_\mathrm{g}=V_\mathrm{g,br}=300$ V) contrast ratio achievable with p-doped Si can be as high as 225 \%, in contrast to 190 \% with ITO (Fig. \ref{fig:Figure4}). This increase in contrast ratio arises because, in depletion mode and for small carrier densities, p-doped Si on SiC suppresses near-field heat transfer ($Q<Q_\mathrm{SiC}$), in contrast to ITO. This suppression in near-field heat transfer stems from the strong temperature dependence of the dielectric function of Si \cite{Fu_DopedSiDrude}. Specifically, when the dielectric functions of the hot and cold sides differ considerably, the thermally excited SPPs from the two sides do not couple efficiently. For instance, there exist regimes of frequencies and carrier densities for which the dielectric functions of p-Si at temperatures $T_\mathrm{1}$ and $T_\mathrm{2}$ may be opposite in sign, hence the SPPs thermal channel is eliminated. Exploiting this effect, the contrast ratio (ON/OFF) of near-field heat transfer can be further enhanced by spectrally misaligning the surface plasmonic resonances of the cold and hot sides of the thermal MOS (Fig. \ref{fig:Figure1}).}

\par{In Fig. \ref{fig:Figure5}b we display the spectral heat flux and corresponding photon transmission probability $\xi$ for points A and B, corresponding to the accumulation and depletion modes in Fig. \ref{fig:Figure5}a, respectively. Similar to the ITO system discussed previously, we note that the large plasmonic losses ($\gamma_\mathrm{p}$) of Si, namely $\gamma_\mathrm{Si}=[26-223]\gamma_\mathrm{SiC}$ for the range of temperatures and carrier densities considered here \cite{Fu_DopedSiDrude} strongly perturb the dispersion of the SiC SPhP when interacting with the p-Si SPP, which makes them hard to distinguish in the plot for $\xi$ in the accumulation mode (right panel).}

\subsection{\label{Graphene}Graphene}

\par{Finally, we consider the system consisting of a graphene sheet on SiC, in the configuration shown in the inset of Fig. \ref{fig:Figure6}a, where we account for the monolayer by setting the thickness of graphene to $L_\mathrm{g}=0.33$ nm, which is the interlayer separation distance of graphite \cite{Zhao_graphenePRB, Zhao_grapheneJHT, Papadakis_graphene}. For the sake of comparison with the ITO and doped Si cases examined previously, we set the vacuum gap thickness to $d=16$ nm in order to maintain the same distance between the SiC interfaces for all three examined materials. We model graphene with the Drude model, by first computing its optical conductivity, which connects to permittivity via $\epsilon_\mathrm{g}(\omega)=1+i\sigma/(\epsilon_\mathrm{o}\omega L_\mathrm{g})$, while taking into account the temperature dependence of both interband and intraband contributions \cite{Falkovsky, Zhao_graphenePRB}. The charge carrier density $N_\mathrm{graphene}$ and Fermi level $E_\mathrm{F}$ are connected via $E_\mathrm{F}=h|v_\mathrm{F}|\sqrt{\pi N_\mathrm{graphene}}$ \cite{Ilic_Graphene2012,Papadakis_graphene}, where $v_\mathrm{F}$ is the Fermi velocity. We take into account the effect of gating on the Fermi velocity and momentum-relaxation rate $\gamma_\mathrm{p}$, following \cite{Papadakis_graphene}.}

\par{In determining the maximum achievable doping and corresponding Fermi level in graphene, the charge per area, $q_\mathrm{A}$ in Eq. \ref{eq:8}, is given by $q_\mathrm{A}=eN_\mathrm{graphene}$, where $N_\mathrm{graphene}$ has units of $\mathrm{m}^{-2}$. The minimum considered carrier density in graphene is $N_\mathrm{graphene}=0$, which corresponds to Fermi level $E_\mathrm{F}=0$ eV, and to $V_\mathrm{g}=0$ V. Extending the range of considered applied bias to $V_\mathrm{g}=[-V_\mathrm{g,br},0]$, i.e. inverting the role of electrons and holes, would yield the same near-field heat transfer ($Q(V_\mathrm{g})$) as for $V_\mathrm{g}>0$, due to linear electronic dispersion of graphene based on which electrons and holes affect its optical properties on equal footing.}

 \begin{figure}[]
\centering
 \includegraphics[width=1\linewidth]{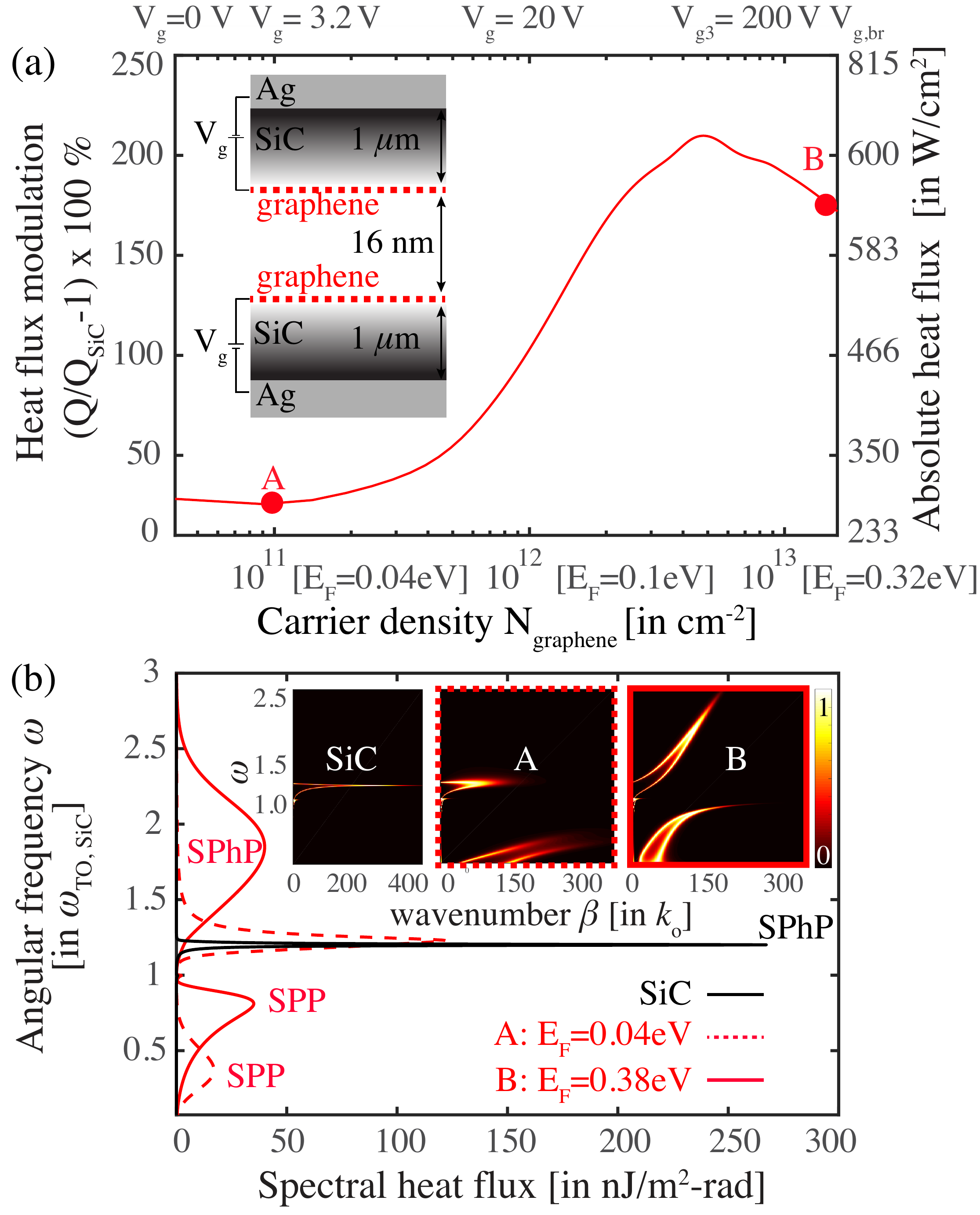}
  \caption{\textbf{Thermal MOS switch with graphene.} (a) Total heat transfer coefficient for $T_\mathrm{1}=600$ K, $T_\mathrm{2}=300$ K, relative to the SiC benchmark, with respect to the carrier density $N_\mathrm{graphene}$ and corresponding Fermi level $E_\mathrm{F}$. Top horizontal axis is the required applied bias $V_\mathrm{g}$ for obtaining the corresponding heat flux. The maximum considered carrier density corresponds to the breakdown condition of SiC, i.e. $V_\mathrm{g,br}=300$ V. The blackbody limit lies at $(Q/Q_\mathrm{SiC}-1)\times 100 \%=-99.7$\%. (b) Spectral heat flux for carrier densities corresponding to points A and B in panel (a), shown with dashed and solid red curves, respectively. Corresponding photon transmission probabilities are shown in the insets. Results are compared to bulk SiC (black curve and left-most $\xi$ curve).}
 \label{fig:Figure6}
\end{figure}

\par{In Fig. \ref{fig:Figure6}a we show the total heat transfer, $Q$, normalized to SiC, as a function of the carrier density. The brackets in the horizontal axis indicate the corresponding Fermi level, $E_\mathrm{F}$, with respect to the charge neutrality point ($N_\mathrm{graphene}=0$). The top horizontal axis provides estimates for the gate voltage $V_\mathrm{g}$ required for obtaining the corresponding near-field heat transfer rate. With respect to bulk SiC, the addition of a monolayer sheet of graphene with tunable carrier density yields a near-field heat transfer modulation that varies between 25 \% and 200 \%. The maximum contrast (ON/OFF) ratio is 147 \%, and maximum heat transfer occurs at a Fermi level of approximately 0.2 eV. Fig. \ref{fig:Figure6}b displays the spectral heat flux corresponding to points A (dashed red curve) and B (solid red curve) in panel (a), i.e. for low- and high-$E_\mathrm{F}$, respectively, in addition to the spectral heat flux for bulk SiC (black curve). At $E_\mathrm{F}=0.04$ eV, we observe two peaks in spectral heat flux, one arising from the SiC SPhP, which lies at $\omega=1.2\omega_\mathrm{TO,SiC}$, and one due to graphene's SPP, near $\omega=0.2\omega_\mathrm{TO,SiC}$, while the corresponding photon transmission probability is shown in the middle inset. The slight blueshift of the SPhP resonance with respect to the bulk SiC case (black line) results from band repulsion, discussed in detail in Fig. \ref{fig:Figure3}. Increasing the Fermi level to $E_\mathrm{F}=0.38$ eV broadens the SPhP peak due to band flatting, as is shown with the photon transmission probability in the right-most inset. Compared to the cases of ITO (Fig. \ref{fig:Figure4}) and p-doped Si (Fig. \ref{fig:Figure5}), plasmonic losses in graphene are reduced by two orders of magnitude \cite{Papadakis_graphene, Zhao_graphenePRB, Zhao_graphenePRB, Falkovsky, Ilic_Graphene2012}, making the graphene SPP mode easily distinguishable from the SiC SPhP mode in the plots of photon transmission probability shown in the insets in Fig. \ref{fig:Figure6}b.}

\section{\label{Conclusions}CONCLUSIONS}

In conclusion, we proposed a thermal MOS switch for active modulation of heat flux in the near-field, as a counterpart to the electronic MOS transistor. The proposed idea stems from the same physical principle of operation as the field effect transistor, namely the accumulation and depletion of charge carriers in an active material electrostatically gated against a back electrode. The near-field heat transfer modulation speed is, therefore, the same as the modulation rate of the field effect, which lies in the GHz range \cite{Speed_fieldeffect1, Speed_fieldeffect2}, and the proposed module can be compatible with current CMOS technology. By performing photonic calculations of near-field heat transfer, while taking into account electrostatic considerations, we showed that by accumulation and depletion of charge carriers in ultra-thin layers of ITO and doped Si, and in monolayer graphene, on SiC, contrast ratios as large as $225\%$ may be achieved. The required gate voltage for modulation is determined by the breakdown field of the material that serves as the oxide in the MOS scheme. SiC is optimal for tunable near-field heat transfer via electrostatic gating due to its good insulating properties, in addition to its pronounced SPhP resonance that optimally transfers heat in the near-field at moderate temperatures. Our results demonstrate the possibility of actively controlling heat flux in the near-field, in a platform adoptable in modern optoelectronics, hence targeting applications in thermal cooling, recycling, and thermal circuitry. 

\section{\label{Methods}Methods}
\par{The computational package used for near-field heat transfer calculations can be found in \cite{CHEN2018163}.}

\section{\label{Authors}Author Information} 
\noindent \textbf{Corresponding author}\\
shanhui@stanford.edu

\noindent \textbf{ORCID}\\
Georgia T. Papadakis: 0000-0001-8107-9221\\
Bo Zhao: 0000-0002-3648-6183\\
Shanhui Fan: 0000-0003-3478-6630\\
 
\noindent \textbf{Notes}\\
The authors declare no competing financial interest.

\section{\label{Acknowledgements}ACKNOWLEDGEMENTS}
\noindent The authors acknowledge the support from the Department of Energy ``Light-Material Interactions in Energy Conversion'' Energy Frontier Research Center under Grant No. DE-SC0001293, and the Department of Energy “Photonics at Thermodynamic Limits” Energy Frontier Research Center under Grant No. DE-SC0019140. G.T. P. acknowledges the TomKat Postdoctoral Fellowship in Sustainable Energy at Stanford University. We acknowledge fruitful discussions with I. Williamson.}

\bibliographystyle{apsrev4-1}
  
%

\end{document}